\documentclass[sigconf]{acmart}
\usepackage{balance}
\usepackage{multirow}
\usepackage{subfig}

 \AtBeginDocument{%
 }

\copyrightyear{2023}
\acmYear{2023}
\setcopyright{rightsretained}
\acmConference[MM '23]{Proceedings of the 31st ACM International Conference on Multimedia}{October 29-November 3, 2023}{Ottawa, ON, Canada}
\acmBooktitle{Proceedings of the 31st ACM International Conference on Multimedia (MM '23), October 29-November 3, 2023, Ottawa, ON, Canada}
\acmDOI{10.1145/3581783.3612848}
\acmISBN{979-8-4007-0108-5/23/10}

\makeatother

\begin{document}

\title{Advancing Audio Emotion and Intent Recognition with Large Pre-Trained Models and Bayesian Inference}



\author{Dejan Porjazovski}
\email{dejan.porjazovski@aalto.fi}
\authornotemark[1]
\affiliation{%
  \institution{Aalto University}
  \city{Espoo}
  \country{Finland}
}

\author{Yaroslav Getman}
\authornote{Both authors contributed equally to this research.}
\email{yaroslav.getman@aalto.fi}
\affiliation{%
  \institution{Aalto University}
  \city{Espoo}
  \country{Finland}
}

\author{Tam\'as Gr\'osz}
\email{tamas.grosz@aalto.fi}
\affiliation{%
  \institution{Aalto University}
  \city{Espoo}
  \country{Finland}
}

\author{Mikko Kurimo}
\email{mikko.kurimo@aalto.fi}
\affiliation{%
  \institution{Aalto University}
  \city{Espoo}
  \country{Finland}
}


\renewcommand{\shortauthors}{Dejan Porjazovski, Yaroslav Getman, Tamás Grósz, \& Mikko Kurimo}

\begin{abstract}
Large pre-trained models are essential in paralinguistic systems, demonstrating effectiveness in tasks like emotion recognition and stuttering detection. In this paper, we employ large pre-trained models for the ACM Multimedia Computational Paralinguistics Challenge, addressing the Requests and Emotion Share tasks. We explore audio-only and hybrid solutions leveraging audio and text modalities. Our empirical results consistently show the superiority of the hybrid approaches over the audio-only models. Moreover, we introduce a Bayesian layer as an alternative to the standard linear output layer. The multimodal fusion approach achieves an 85.4\% UAR on HC-Requests and 60.2\% on HC-Complaints. The ensemble model for the Emotion Share task yields the best $\rho$ value of .614. The Bayesian wav2vec2 approach, explored in this study, allows us to easily build ensembles, at the cost of fine-tuning only one model. Moreover, we can have usable confidence values instead of the usual overconfident posterior probabilities.
\end{abstract}




\maketitle

\section{Introduction}
In the era of voice-based human-computer interaction devices, the significance of paralinguistics as an integral component cannot be ignored. Paralinguistics is a field dedicated to studying various traits of the speaker. As such, it has gained importance in ensuring effective communication between humans and machines. As these voice-based systems continue to develop, understanding and interpreting paralinguistic cues such as tone, pitch, rhythm, and emotional expressions has become crucial for enhancing the user experience and enabling more natural and intuitive interactions. This year, the ACM Multimedia Computational Paralinguistics Challenge (ComParE 2023) introduced two paralinguistic tasks \cite{Schuller23-TAM}, Requests, utilising the HealthCall30 corpus (HC-C) \cite{lackovic2022prediction} and Emotion Share, utilising the Hume-Prosody dataset (HP-C) \cite{cowen2019mapping}.



When addressing paralinguistic tasks, a popular approach is to employ features extracted from pre-trained models \cite{sheikh2022introducing}. The organisers of this year's competition presented several solutions as baselines such as DeepSpectrum \cite{amiriparian2017Snore}, AuDeep \cite{Amiriparian2017SequenceTS, freitag2017audeep} and the ComParE Acoustic Feature Set. 
Lastly, the popular pre-trained wav2vec2 model \cite{baevski2020wav2vec}, which has exhibited remarkable results in various paralinguistic domains \cite{pepino21_interspeech, grosz2022wav2vec2, shor22_interspeech, getman2022wav2vec2, vaessen2022fine}, was also employed as a baseline. 



This study follows the trend of leveraging large pre-trained models, particularly wav2vec2 variants, which are trained on extensive audio data and suitable for tasks with limited in-domain data. We experiment with audio-only and hybrid solutions that combine audio and text modalities. What sets our work apart from the previous studies is that we aimed to build a model that is able to signal its confidence. Standard, fine-tuned models are usually overconfident in their prediction, and their posteriors are not applicable to assess their confidence level. By connecting a Bayesian output layer~\cite{Goan2020} instead of the traditional linear layer, we can easily create an ensemble of fine-tuned models at the cost of training only one large network. Moreover, incorporating the Bayesian layer allows us to measure the uncertainty associated with the predictions, providing valuable insights into the decision-making process.

\section{Methods and experiments}

As discussed earlier, using pre-trained models is a popular choice for solving paralinguistic tasks. To this end, for extracting audio features, we experimented with fine-tuning various wav2vec2 models. Moreover, we investigated the significance of different layers for the task, as an alternative to the commonly used approach of averaging or using the last Transformer layer. For generating predictions, apart from the standard linear layer, we additionally experimented with a Bayesian linear layer.

In Bayesian neural networks \cite{Goan2020}, instead of learning single parameter values, the goal is to optimise their posterior distribution $p(w|D)$. To compute the posterior, we need to calculate the data likelihood. The data likelihood requires integrating over all the possible weights, which is intractable. Instead of calculating the exact posterior, we can approximate it using Stochastic Variational Inference (SVI) \cite{hoffman2013stochastic}. The SVI defines a parameterised approximate posterior $p_\phi(w)$, where the parameters $\phi$ are used to tune the approximate posterior closer to the original one. To bring the approximate posterior closer to the exact one, we need a way to measure the dissimilarity between the distributions. One way to measure the dissimilarity is by using the Kullback-Leibler (KL) divergence. However, estimating the KL divergence would still require the original posterior $p(w|D)$, which is intractable. Luckily, from the KL divergence, a tractable solution can be derived, called evidence lower bound (ELBO)~\cite{sun2018functional}, used as an objective function for the SVI.


The Requests task falls within the broader field of intent recognition, which has seen a shift towards end-to-end solutions \cite{desot2019towards, serdyuk2018towards}. However, these solutions still lag behind the pipeline approach, which involves generating transcripts using an automatic speech recognition (ASR) system and then using those transcripts as input for a text-based classifier. Motivated by this, we extracted text-based features using a pre-trained BERT \cite{devlin-etal-2019-bert} model. Since both the Requests and Emotion Share tasks come with only audio files, we used the Whisper model \cite{radford2022robust}, which produces transcripts, preserving the capitalisation and punctuation. To combine the audio and text modalities, we experimented with early and late fusion techniques.


We implemented all the models using the PyTorch framework \cite{paszke2019pytorch}. Some experiments utilised the SpeechBrain toolkit \cite{ravanelli2021speechbrain}, while the Bayesian experiments employed the implementation described in \cite{lee2022graddiv}. The negative log-likelihood was optimised for the Requests task, and the mean squared error (MSE) was used for Emotions. For a detailed implementation description and a comprehensive list of hyperparameters, please refer to the code repository\footnote{https://github.com/aalto-speech/ComParE2023}.

\subsection{Requests task}
The Requests task involves a classification problem that can be further divided into two binary sub-tasks: HC-Complaints and HC-Requests. Consequently, we have two options for modelling this task: treating it as a combined 4-class classification problem or developing separate binary classification models. For the combined 4-class classification, we merged the HC-Complaints and HC-Requests labels in pairs, resulting in the following classes: "no\_affil", "yes\_affil", "no\_presta", and "yes\_presta".


For audio feature extraction, we used the multilingual French ASR wav2vec2 model (wav2vec2-large-xlsr-53-french), which we trained by updating both the convolutional feature encoder and the contextual Transformer layers. 
The priors for the Bayesian layer were set as the mean and standard deviation of the learned weights from the model trained with the standard linear output layer. To determine the optimal layer for each task, we conducted a layer-wise analysis by fine-tuning the model on a subset of the data. Based on those findings, we used the specific layers in the subsequent experiments.

To generate the transcripts, we used the large Whisper version 2 model (whisper-large-v2), trained on 680k hours of labelled data. To extract the text features, we used the French BERT model \cite{martin-etal-2020-camembert} (camembert-base). For combining both modalities, we employed the weighted late fusion approach, which calculates the weighted sum of the probabilities with weights tuned using the development set: 1.0/0.9 and 1.0/0.5 (BERT/wav2vec2) for the HC-Requests and HC-Complaints, respectively.

\subsection{Emotion Share task}
To extract the audio features for the Emotion Share task, we experimented with two pre-trained wav2vec2 models. The first one is the large pre-trained multilingual XLSR model \cite{conneau21_interspeech} (wav2vec2-large-xlsr-53). In the initial experiments, we chose this model due to the multilingual nature of the data. As a second choice, we employed the English pre-trained and ASR fine-tuned version (wav2vec2-large-960h-lv60-self). The justification for this model is due to the speakers in the dataset being from the United States, South Africa, and Venezuela, meaning that in the majority of the cases, the English language is being spoken. During training, we optimised both the convolutional feature encoder and the contextual Transformer layers, following a similar approach as the Requests task. To generate the emotion intensities, we applied a sigmoid function to the logits produced by either the standard linear or the Bayesian linear layer. This normalisation was necessary because the emotion intensities in the dataset were scaled between 0 and 1. Similar to the Requests task, the priors for the mean and standard deviation were taken from the standard linear layer. To investigate the impact that the different layers have, we performed a layer-wise analysis by fine-tuning the model on a subset of the data. The preliminary experiments revealed that layer 18 was the most optimal for the multilingual model and the last layer for the English.

Transcripts for this task were also generated using the large Whisper model, and the features were extracted using the [CLS] token of the large cased English BERT model (bert-large-cased). To combine the audio and text modalities, we adopted early and late fusion approaches. 
In the early fusion, the wav2vec2 and BERT models were trained separately. Then, the embeddings from both modalities were concatenated. Finally, a separate multi-output regression model was trained using the concatenated embeddings.


\section{Results and analysis}
\subsection{Requests}
The results of the experiments conducted on the Requests task are summarised in Table \ref{tab:res_req_submissions}. To select the best candidates for the submissions, we conducted evaluations using multiple approaches on the development set. The initial focus was on the 4-class classification problem, where we trained the first two models (models 1 and 2). The results revealed that the Bayesian output layer slightly outperformed the standard linear model on the HC-Requests sub-task while showcasing a more substantial advantage on the HC-Complaints. Next, we trained separate models for each sub-task, which demonstrated a significant improvement compared to the 4-class approach (models 3 and 4). In this case, the standard linear model performed slightly better than the Bayesian on the HC-Requests sub-task, while the Bayesian exhibited a slight advantage on the HC-Complaints sub-task.

\begin{table}[t]
    \caption{UARs on the Requests task. The layer column indicates the number of Transformer blocs used for the Requests and Complaints sub-tasks, respectively.}
  \label{tab:res_req_submissions}
  \begin{tabular}{l|l|l|cc|cc}
    \toprule
    \multirow{2}{*}{Model} & \multirow{2}{*}{Task} & \multirow{2}{*}{Layer} & \multicolumn{2}{c|}{Requests} & \multicolumn{2}{c}{Complaints} \\
    & & & Dev & Test & Dev & Test \\
    \midrule
    1. $wav2vec2_{L}$ & \multirow{2}{*}{\begin{tabular}[c]{@{}l@{}}4 classes\\ audio\end{tabular}} & 19  & 77.0 & / & 52.3 & / \\
    2. $wav2vec2_{B}$ & & 19 & 77.7 & 80.8 & 56.8 & 58.6 \\
    \cmidrule(lr){1-7}
    3. $wav2vec2_{L}$ & \multirow{2}{*}{\begin{tabular}[c]{@{}l@{}}2 classes\\ audio\end{tabular}} & 18/12 & 79.3 & 82.2 & 58.4 & 57.7 \\
    4. $wav2vec2_{B}$ &  & 18/12 & 78.7 & 82.1 & 59.1 & 58.8 \\
    \cmidrule(lr){1-7}
    5. $BERT_{L}$ & \multirow{2}{*}{\begin{tabular}[c]{@{}l@{}}2 classes\\ text\end{tabular}} & CLS & 79.5 & / & 64.0 & / \\
    6. $BERT_{B}$ & & CLS & 80.1 & / & 63.2 & / \\
    \cmidrule(lr){1-7}
    7. \begin{tabular}[c]{@{}l@{}} $wav2vec2_L$ \\ $ + BERT$ LF \end{tabular} & \begin{tabular}[c]{@{}l@{}} 2 classes \\ multimodal\end{tabular} & 18/12 & \textbf{81.7} & \textbf{85.4} & \textbf{64.9} & \textbf{60.2} \\
    \cmidrule(lr){1-7}
    \begin{tabular}[c]{@{}l@{}} 8. Ensemble of \\ 2, 3, 4, and 7 \end{tabular} & / & / & / & 84.2 & / & 59.6 \\
    \cmidrule(lr){1-7}
    wav2vec2 \cite{Schuller23-TAM} & baseline & avg & 65.1 & 67.2 & 50.9 & 52.2 \\
  \bottomrule
\end{tabular}
\end{table}


Continuing our experiments, we sought to compare text-based solutions with the audio-only models. Since the 2-class training yielded better overall results, we utilised the same setting for the text-based systems. Interestingly, unlike the audio models, the standard linear model demonstrated slightly better performance on the HC-Complaints sub-task, while the Bayesian model showed superiority on the HC-Requests sub-task. Moreover, the text-based solutions demonstrated improvement over the audio-only models for both sub-tasks. Notably, we observed a 0.8\% absolute improvement on the development set over the best audio model on the HC-Requests sub-task and 4.9\% on the HC-Complaints.

In the final experiment, we explored the benefits of combining both audio and text modalities using the late fusion approach (model 7). This fusion of modalities resulted in the best overall performance on the development set, highlighting the advantages of leveraging both audio and text information.

Overall, the aforementioned approaches outperformed the wav2vec2 baseline system on the development set, with the best model (model 7) achieving a 16.6\% absolute UAR improvement on the HC-Requests and a 14\% on the HC-Complaints sub-task.

The preliminary experiments conducted on the development set provided us with promising candidates for the final submissions. Notably, for the final submissions, we re-trained the models using the train and development sets.

Our first selected candidate was the model trained with 4 classes using a Bayesian linear layer (model 2). We chose this model based on its superior performance compared to the model with a standard linear layer. Additionally, we wanted to investigate the potential benefits of training a single model with 4 classes, instead of two separate ones. This 4-class Bayesian model achieved 80.8\% UAR score on the HC-Requests and 58.6\% on the HC-Complaints, outperforming the wav2vec2 baseline.

As the following two candidates, we opted for the standard linear and Bayesian linear models trained on audio-only data using 2-class classification (models 3 and 4). The purpose was to compare the performance of the Bayesian and standard linear layers. The results showed that the model with the Bayesian layer achieved overall better performance. Furthermore, the 2-class Bayesian model outperformed its 4-class counterpart, showcasing the benefit of separate models for each sub-task.

The fusion of audio and text modalities (model 7) yielded the best results on the development set. Hence, we selected this hybrid model as our fourth submission to be evaluated on the test set. As evident from the results, this model achieved superior performance on both the HC-Requests and HC-Complaints sub-tasks, compared to the other submissions.

For our last submission, we created an ensemble of the four selected models by utilising majority voting to determine the final predictions. Although this model combined all previous submissions, it fell slightly behind the multimodal approach.


\subsection{Emotion Share}
\begin{figure*}[h!]
\centering
  \subfloat[PDF on the HC-Complaints]{\includegraphics[width=190pt, height=80pt]{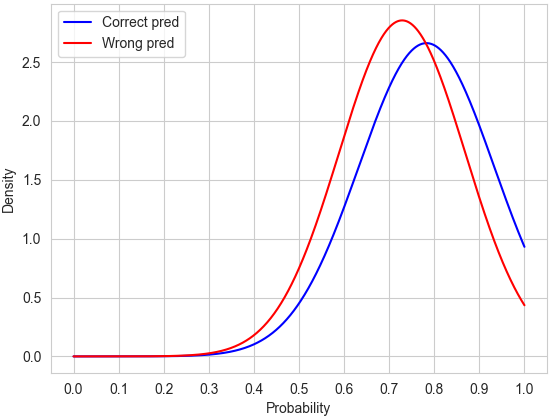}\label{fig:gaussian_complaints}}
  \hspace{1.5cm}
  \subfloat[PDF on the HC-Requests.]{\includegraphics[width=190pt, height=80pt]{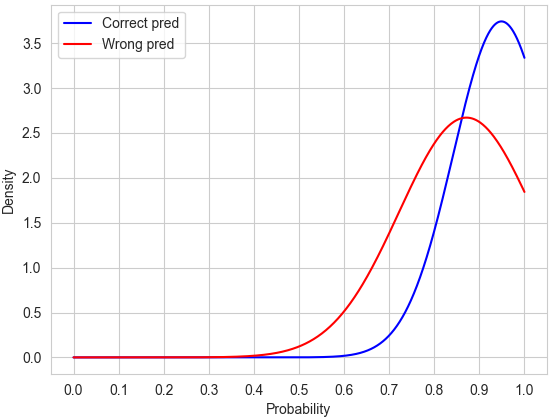}\label{fig:gaussian_requests}}
  \caption{PDF for the correct and wrong predictions on the HC-Complaints and HC-Requests sub-tasks.}
  \label{fig:gaussian_pdf}
\end{figure*}

Following a similar approach to the Requests task, to select the most suitable candidates for submission, we conducted preliminary experiments on the development set. A summary of our experiments can be found in Table \ref{tab:res_emo_submissions}.

\begin{table}[t]
  \caption{Spearman's $\rho$ for the dev and test sets on the Emotion Share task.}
  \label{tab:res_emo_submissions}
  \begin{tabular}{l|l|cc}
    \toprule
    Model & Layer & $\rho$ Dev & $\rho$ Test \\
    \midrule
    1. $wav2vec2_{L}^{ml}$  segment-based & 18  & .568 & .543 \\
    2. $wav2vec2_{B}^{ml}$  segment-based & 18 & .502 & / \\
    \cmidrule(lr){1-4}
    3. $wav2vec2_{L}^{en}$ & last & .566 & .587 \\
    4. $wav2vec2_{B}^{en}$ & last & .561 & .592 \\
    \cmidrule(lr){1-4}
    5. $wav2vec2_{L}^{en}+BERT$ Early Fusion & last/CLS & .577 & .612 \\
    6. $wav2vec2_{L}^{en}+BERT$ Late Fusion & last/CLS & .571 & / \\
    \cmidrule(lr){1-4}
    7. Ensemble of 3, 4, and 5 & / & / & \textbf{.614} \\
    \midrule
    Baseline (wav2vec2) \cite{Schuller23-TAM} & avg & .500 & .514 \\
  \bottomrule
\end{tabular}
\end{table}



As an initial set of experiments, we explored processing the utterances in segments using overlapping windows. Specifically, we used a segment size of 30ms with a 10ms stride. We chose the segment and stride sizes based on preliminary experiments. To address the multilingual nature of the data, we conducted experiments with both the multilingual and English wav2vec2 models. With the segment-based processing approach, we observed better performance from the multilingual model. Therefore, in the table, we only present the results from that model. Notably, the comparison between models 1 and 2 revealed that the standard linear layer performed substantially better than the Bayesian.

Subsequently, we evaluated the English wav2vec2 model using both standard linear and Bayesian layers (models 3 and 4). The results indicated a slight advantage for the standard linear layer in the Emotion Share task, although the difference was negligible. Additionally, the segment-based processing with the standard linear layer (model 1) showed slightly better results compared to the English model that processed the entire utterance at once. However, due to the Bayesian solution being more stable using the English model, we proceeded with that for the subsequent experiments.

In addition to audio-only techniques, we also explored multimodal solutions by fusing audio and text systems  (models 5 and 6). The results revealed that early fusion slightly outperformed the late fusion approach. Furthermore, both approaches outperformed the audio-only solutions, demonstrating the benefits of leveraging both audio and text modalities. 

After obtaining results on the development set, we picked the most promising models to evaluate on the test set. First, we chose the multilingual model with segment-based processing and a linear classification layer (model~1). This model achieved a $\rho$ score of .543, surpassing the baseline performance.
Next, we compared the English wav2vec2 models using both the standard linear and Bayesian layers (models 3 and 4). The results demonstrated that the Bayesian approach outperformed the standard linear alternative. This is consistent with the findings in the Requests task. Moreover, both approaches yielded better performance than the segment-based processing solution.

To assess the effectiveness of modality fusion, we selected the hybrid multimodal model with early fusion (model 5) for our next submission. Based on the results obtained on the test set, we concluded its superiority over the audio-only solutions.

For our final submission, we opted again for an ensemble approach. However, instead of combining all previous submissions, we excluded model 1 due to its inferior results compared to the others. The ensemble was created by averaging the intensities of each model's predictions, resulting in the best performance and surpassing the baseline by a $\rho$ value of .1.


\subsection{Measuring model uncertainty}
One big advantage of Bayesian neural networks is their ability to model the uncertainty that comes with the predictions. This brings us a step closer to understanding the reasons behind the predictions, which is more difficult to achieve using standard networks.


To delve into the decision-making process, we plotted the probability density function (PDF) for the probabilities of the correct and wrong predictions, presented in Figure \ref{fig:gaussian_pdf}. To conduct the experiments, we used the Bayesian wav2vec2 model (model 4 from Table \ref{tab:res_req_submissions}). Since Bayesian models allow for sampling different weights, we sampled 500 of them, resulting in that many probabilities per sample. Then, we averaged the probabilities to get one probability per sample. By looking at the density for the HC-Complaints sub-task (Figure \ref{fig:gaussian_complaints}), we can notice that the mean of the incorrect predictions is smaller than the mean of the correct ones, indicating that when the model is wrong, it is also less confident about the prediction. A similar trend can be observed for the HC-Requests sub-task (Figure \ref{fig:gaussian_requests}), where the mean of the correct predictions is bigger than the average of the wrong ones. Additionally, the erroneous predictions have higher variance, indicating higher uncertainty.

\section{Conclusions}

This study addressed the Requests and Emotion Share challenges within the ACM Multimedia ComParE challenge. Our findings highlighted the advantages of incorporating both audio and text modalities in a hybrid approach, surpassing the performance of the audio-only solutions. Through our exploration of different wav2vec2 models and layers, we demonstrated the importance of carefully selecting the appropriate settings for each specific task.



Our multimodal late fusion approach achieved UAR scores of 85.4\% and 60.2\% for the HC-Requests and HC-Complaints sub-tasks, respectively, corresponding to absolute improvements of 18.2\% and 8\% compared to the wav2vec2 baseline. Notably, the consistency between the development and test results across all models indicates that they did not overfit, further validating their effectiveness. Similar improvements were observed in the Emotion Share task, where our best-performing model, utilising an early fusion approach, yielded a notable $\rho$ score of .612. Furthermore, an ensemble of our top-performing models resulted in a slight improvement, achieving a $\rho$ score of .614. 

By incorporating the Bayesian layer, we observed improvements over the standard linear layer. Moreover, we gained deeper insights into the decision-making process of our models. Through PDF analysis, we observed that the model exhibited reduced confidence in its predictions while making errors. In the future, this property could be exploited to develop systems that could accurately inform users about their confidence and possible mistakes.


\begin{acks}
This work was supported by the Foundation for Aalto University Science and Technology. We are grateful for the Academy of Finland project funding number 337073 and 345790. The computational resources were provided by Aalto ScienceIT.
\end{acks}

\bibliographystyle{ACM-Reference-Format}
\balance
\bibliography{references}


\begin{thebibliography}{25}


\ifx \showCODEN    \undefined \def \showCODEN     #1{\unskip}     \fi
\ifx \showDOI      \undefined \def \showDOI       #1{#1}\fi
\ifx \showISBNx    \undefined \def \showISBNx     #1{\unskip}     \fi
\ifx \showISBNxiii \undefined \def \showISBNxiii  #1{\unskip}     \fi
\ifx \showISSN     \undefined \def \showISSN      #1{\unskip}     \fi
\ifx \showLCCN     \undefined \def \showLCCN      #1{\unskip}     \fi
\ifx \shownote     \undefined \def \shownote      #1{#1}          \fi
\ifx \showarticletitle \undefined \def \showarticletitle #1{#1}   \fi
\ifx \showURL      \undefined \def \showURL       {\relax}        \fi
\providecommand\bibfield[2]{#2}
\providecommand\bibinfo[2]{#2}
\providecommand\natexlab[1]{#1}
\providecommand\showeprint[2][]{arXiv:#2}

\bibitem[Amiriparian et~al\mbox{.}(2017a)]%
        {Amiriparian2017SequenceTS}
\bibfield{author}{\bibinfo{person}{Shahin Amiriparian}, \bibinfo{person}{Michael~J. Freitag}, \bibinfo{person}{Nicholas Cummins}, {and} \bibinfo{person}{Bj{\"o}rn Schuller}.} \bibinfo{year}{2017}\natexlab{a}.
\newblock \showarticletitle{Sequence to Sequence Autoencoders for Unsupervised Representation Learning from Audio}. In \bibinfo{booktitle}{\emph{Workshop on Detection and Classification of Acoustic Scenes and Events}}.
\newblock


\bibitem[Amiriparian et~al\mbox{.}(2017b)]%
        {amiriparian2017Snore}
\bibfield{author}{\bibinfo{person}{Shahin Amiriparian}, \bibinfo{person}{Maurice Gerczuk}, \bibinfo{person}{Sandra Ottl}, \bibinfo{person}{Nicholas Cummins}, \bibinfo{person}{Michael Freitag}, \bibinfo{person}{Sergey Pugachevskiy}, \bibinfo{person}{Alice Baird}, {and} \bibinfo{person}{Bj{\"o}rn Schuller}.} \bibinfo{year}{2017}\natexlab{b}.
\newblock \showarticletitle{Snore Sound Classification Using Image-Based Deep Spectrum Features}. In \bibinfo{booktitle}{\emph{Interspeech 2017}}. \bibinfo{publisher}{{ISCA}}, \bibinfo{pages}{3512--3516}.
\newblock


\bibitem[Baevski et~al\mbox{.}(2020)]%
        {baevski2020wav2vec}
\bibfield{author}{\bibinfo{person}{Alexei Baevski}, \bibinfo{person}{Yuhao Zhou}, \bibinfo{person}{Abdelrahman Mohamed}, {and} \bibinfo{person}{Michael Auli}.} \bibinfo{year}{2020}\natexlab{}.
\newblock \showarticletitle{wav2vec 2.0: A framework for self-supervised learning of speech representations}.
\newblock \bibinfo{journal}{\emph{Advances in neural information processing systems}}  \bibinfo{volume}{33} (\bibinfo{year}{2020}), \bibinfo{pages}{12449--12460}.
\newblock


\bibitem[Conneau et~al\mbox{.}(2021)]%
        {conneau21_interspeech}
\bibfield{author}{\bibinfo{person}{Alexis Conneau}, \bibinfo{person}{Alexei Baevski}, \bibinfo{person}{Ronan Collobert}, \bibinfo{person}{Abdelrahman Mohamed}, {and} \bibinfo{person}{Michael Auli}.} \bibinfo{year}{2021}\natexlab{}.
\newblock \showarticletitle{{Unsupervised Cross-Lingual Representation Learning for Speech Recognition}}. In \bibinfo{booktitle}{\emph{Interspeech 2021}}. \bibinfo{pages}{2426--2430}.
\newblock
\urldef\tempurl%
\url{https://doi.org/10.21437/Interspeech.2021-329}
\showDOI{\tempurl}


\bibitem[Cowen et~al\mbox{.}(2019)]%
        {cowen2019mapping}
\bibfield{author}{\bibinfo{person}{Alan~S Cowen}, \bibinfo{person}{Hillary~Anger Elfenbein}, \bibinfo{person}{Petri Laukka}, {and} \bibinfo{person}{Dacher Keltner}.} \bibinfo{year}{2019}\natexlab{}.
\newblock \showarticletitle{Mapping 24 emotions conveyed by brief human vocalization.}
\newblock \bibinfo{journal}{\emph{American Psychologist}} \bibinfo{volume}{74}, \bibinfo{number}{6} (\bibinfo{year}{2019}), \bibinfo{pages}{698}.
\newblock


\bibitem[Desot et~al\mbox{.}(2019)]%
        {desot2019towards}
\bibfield{author}{\bibinfo{person}{Thierry Desot}, \bibinfo{person}{Fran{\c{c}}ois Portet}, {and} \bibinfo{person}{Michel Vacher}.} \bibinfo{year}{2019}\natexlab{}.
\newblock \showarticletitle{Towards end-to-end spoken intent recognition in smart home}. In \bibinfo{booktitle}{\emph{2019 International Conference on Speech Technology and Human-Computer Dialogue (SpeD)}}. IEEE, \bibinfo{pages}{1--8}.
\newblock


\bibitem[Devlin et~al\mbox{.}(2019)]%
        {devlin-etal-2019-bert}
\bibfield{author}{\bibinfo{person}{Jacob Devlin}, \bibinfo{person}{Ming-Wei Chang}, \bibinfo{person}{Kenton Lee}, {and} \bibinfo{person}{Kristina Toutanova}.} \bibinfo{year}{2019}\natexlab{}.
\newblock \showarticletitle{{BERT}: Pre-training of Deep Bidirectional Transformers for Language Understanding}. In \bibinfo{booktitle}{\emph{NAACL: Human Language Technologies, Volume 1 (Long and Short Papers)}}. \bibinfo{publisher}{Association for Computational Linguistics}, \bibinfo{address}{Minneapolis, Minnesota}.
\newblock
\urldef\tempurl%
\url{https://doi.org/10.18653/v1/N19-1423}
\showDOI{\tempurl}


\bibitem[Freitag et~al\mbox{.}(2017)]%
        {freitag2017audeep}
\bibfield{author}{\bibinfo{person}{Michael Freitag}, \bibinfo{person}{Shahin Amiriparian}, \bibinfo{person}{Sergey Pugachevskiy}, \bibinfo{person}{Nicholas Cummins}, {and} \bibinfo{person}{Bj{\"o}rn Schuller}.} \bibinfo{year}{2017}\natexlab{}.
\newblock \showarticletitle{audeep: Unsupervised learning of representations from audio with deep recurrent neural networks}.
\newblock \bibinfo{journal}{\emph{The Journal of Machine Learning Research}} \bibinfo{volume}{18}, \bibinfo{number}{1} (\bibinfo{year}{2017}), \bibinfo{pages}{6340--6344}.
\newblock


\bibitem[Getman et~al\mbox{.}(2022)]%
        {getman2022wav2vec2}
\bibfield{author}{\bibinfo{person}{Yaroslav Getman}, \bibinfo{person}{Ragheb Al-Ghezi}, \bibinfo{person}{Katja Voskoboinik}, \bibinfo{person}{Tam{\'a}s Gr{\'o}sz}, \bibinfo{person}{Mikko Kurimo}, \bibinfo{person}{Giampiero Salvi}, \bibinfo{person}{Torbj{\o}rn Svendsen}, {and} \bibinfo{person}{Sofia Str{\"o}mbergsson}.} \bibinfo{year}{2022}\natexlab{}.
\newblock \showarticletitle{Wav2vec2-based speech rating system for children with speech sound disorder}. In \bibinfo{booktitle}{\emph{Interspeech}}.
\newblock


\bibitem[Goan and Fookes(2020)]%
        {Goan2020}
\bibfield{author}{\bibinfo{person}{Ethan Goan} {and} \bibinfo{person}{Clinton Fookes}.} \bibinfo{year}{2020}\natexlab{}.
\newblock \bibinfo{booktitle}{\emph{Bayesian Neural Networks: An Introduction and Survey}}.
\newblock \bibinfo{publisher}{Springer International Publishing}, \bibinfo{address}{Cham}, \bibinfo{pages}{45--87}.
\newblock
\showISBNx{978-3-030-42553-1}
\urldef\tempurl%
\url{https://doi.org/10.1007/978-3-030-42553-1_3}
\showDOI{\tempurl}


\bibitem[Gr{\'o}sz et~al\mbox{.}(2022)]%
        {grosz2022wav2vec2}
\bibfield{author}{\bibinfo{person}{Tam{\'a}s Gr{\'o}sz}, \bibinfo{person}{Dejan Porjazovski}, \bibinfo{person}{Yaroslav Getman}, \bibinfo{person}{Sudarsana Kadiri}, {and} \bibinfo{person}{Mikko Kurimo}.} \bibinfo{year}{2022}\natexlab{}.
\newblock \showarticletitle{Wav2vec2-based paralinguistic systems to recognise vocalised emotions and stuttering}. In \bibinfo{booktitle}{\emph{Proceedings of the 30th ACM International Conference on Multimedia}}. \bibinfo{pages}{7026--7029}.
\newblock


\bibitem[Hoffman et~al\mbox{.}(2013)]%
        {hoffman2013stochastic}
\bibfield{author}{\bibinfo{person}{Matthew~D Hoffman}, \bibinfo{person}{David~M Blei}, \bibinfo{person}{Chong Wang}, {and} \bibinfo{person}{John Paisley}.} \bibinfo{year}{2013}\natexlab{}.
\newblock \showarticletitle{Stochastic variational inference}.
\newblock \bibinfo{journal}{\emph{Journal of Machine Learning Research}} (\bibinfo{year}{2013}).
\newblock


\bibitem[Lackovic et~al\mbox{.}(2022)]%
        {lackovic2022prediction}
\bibfield{author}{\bibinfo{person}{Nikola Lackovic}, \bibinfo{person}{Claude Montaci{\'e}}, \bibinfo{person}{Gauthier Lalande}, {and} \bibinfo{person}{Marie-Jos{\'e} Caraty}.} \bibinfo{year}{2022}\natexlab{}.
\newblock \showarticletitle{Prediction of User Request and Complaint in Spoken Customer-Agent Conversations}.
\newblock \bibinfo{journal}{\emph{arXiv preprint arXiv:2208.10249}} (\bibinfo{year}{2022}).
\newblock


\bibitem[Lee et~al\mbox{.}(2022)]%
        {lee2022graddiv}
\bibfield{author}{\bibinfo{person}{Sungyoon Lee}, \bibinfo{person}{Hoki Kim}, {and} \bibinfo{person}{Jaewook Lee}.} \bibinfo{year}{2022}\natexlab{}.
\newblock \showarticletitle{Graddiv: Adversarial robustness of randomized neural networks via gradient diversity regularization}.
\newblock \bibinfo{journal}{\emph{IEEE Transactions on Pattern Analysis and Machine Intelligence}} (\bibinfo{year}{2022}).
\newblock


\bibitem[Martin et~al\mbox{.}(2020)]%
        {martin-etal-2020-camembert}
\bibfield{author}{\bibinfo{person}{Louis Martin}, \bibinfo{person}{Benjamin Muller}, \bibinfo{person}{Pedro~Javier Ortiz~Su{\'a}rez}, \bibinfo{person}{Yoann Dupont}, \bibinfo{person}{Laurent Romary}, \bibinfo{person}{{\'E}ric de~la Clergerie}, \bibinfo{person}{Djam{\'e} Seddah}, {and} \bibinfo{person}{Beno{\^\i}t Sagot}.} \bibinfo{year}{2020}\natexlab{}.
\newblock \showarticletitle{{C}amem{BERT}: a Tasty {F}rench Language Model}. In \bibinfo{booktitle}{\emph{Proceedings of the 58th Annual Meeting of the Association for Computational Linguistics}}. \bibinfo{publisher}{ACL}, \bibinfo{address}{Online}.
\newblock
\urldef\tempurl%
\url{https://doi.org/10.18653/v1/2020.acl-main.645}
\showDOI{\tempurl}


\bibitem[Paszke et~al\mbox{.}(2019)]%
        {paszke2019pytorch}
\bibfield{author}{\bibinfo{person}{Adam Paszke}, \bibinfo{person}{Sam Gross}, \bibinfo{person}{Francisco Massa}, \bibinfo{person}{Adam Lerer}, \bibinfo{person}{James Bradbury}, \bibinfo{person}{Gregory Chanan}, \bibinfo{person}{Trevor Killeen}, \bibinfo{person}{Zeming Lin}, \bibinfo{person}{Natalia Gimelshein}, \bibinfo{person}{Luca Antiga}, {et~al\mbox{.}}} \bibinfo{year}{2019}\natexlab{}.
\newblock \showarticletitle{Pytorch: An imperative style, high-performance deep learning library}.
\newblock \bibinfo{journal}{\emph{Advances in neural information processing systems}}  \bibinfo{volume}{32} (\bibinfo{year}{2019}).
\newblock


\bibitem[Pepino et~al\mbox{.}(2021)]%
        {pepino21_interspeech}
\bibfield{author}{\bibinfo{person}{Leonardo Pepino}, \bibinfo{person}{Pablo Riera}, {and} \bibinfo{person}{Luciana Ferrer}.} \bibinfo{year}{2021}\natexlab{}.
\newblock \showarticletitle{{Emotion Recognition from Speech Using wav2vec 2.0 Embeddings}}. In \bibinfo{booktitle}{\emph{Interspeech 2021}}. \bibinfo{pages}{3400--3404}.
\newblock
\urldef\tempurl%
\url{https://doi.org/10.21437/Interspeech.2021-703}
\showDOI{\tempurl}


\bibitem[Radford et~al\mbox{.}(2022)]%
        {radford2022robust}
\bibfield{author}{\bibinfo{person}{Alec Radford}, \bibinfo{person}{Jong~Wook Kim}, \bibinfo{person}{Tao Xu}, \bibinfo{person}{Greg Brockman}, \bibinfo{person}{Christine McLeavey}, {and} \bibinfo{person}{Ilya Sutskever}.} \bibinfo{year}{2022}\natexlab{}.
\newblock \showarticletitle{Robust speech recognition via large-scale weak supervision}.
\newblock \bibinfo{journal}{\emph{arXiv preprint arXiv:2212.04356}} (\bibinfo{year}{2022}).
\newblock


\bibitem[Ravanelli et~al\mbox{.}(2021)]%
        {ravanelli2021speechbrain}
\bibfield{author}{\bibinfo{person}{Mirco Ravanelli}, \bibinfo{person}{Titouan Parcollet}, \bibinfo{person}{Peter Plantinga}, \bibinfo{person}{Aku Rouhe}, \bibinfo{person}{Samuele Cornell}, \bibinfo{person}{Loren Lugosch}, \bibinfo{person}{Cem Subakan}, \bibinfo{person}{Nauman Dawalatabad}, \bibinfo{person}{Abdelwahab Heba}, \bibinfo{person}{Jianyuan Zhong}, {et~al\mbox{.}}} \bibinfo{year}{2021}\natexlab{}.
\newblock \showarticletitle{SpeechBrain: A general-purpose speech toolkit}.
\newblock \bibinfo{journal}{\emph{arXiv preprint arXiv:2106.04624}} (\bibinfo{year}{2021}).
\newblock


\bibitem[Schuller et~al\mbox{.}(2023)]%
        {Schuller23-TAM}
\bibfield{author}{\bibinfo{person}{Bj\"{o}rn~W.\ Schuller}, \bibinfo{person}{Anton Batliner}, \bibinfo{person}{Shahin Amiriparian}, \bibinfo{person}{Alexander Barnhill}, \bibinfo{person}{Maurice Gerczuk}, \bibinfo{person}{Andreas Triantafyllopoulos}, \bibinfo{person}{Alice Baird}, \bibinfo{person}{Panagiotis Tzirakis}, \bibinfo{person}{Chris Gagne}, \bibinfo{person}{Alan~S.\ Cowen}, \bibinfo{person}{Nikola Lackovic}, \bibinfo{person}{Marie-Jos\'e Caraty}, {and} \bibinfo{person}{Claude Montaci\'e}.} \bibinfo{year}{2023}\natexlab{}.
\newblock \showarticletitle{{The ACM Multimedia 2023 Computational Paralinguistics Challenge: Emotion Share \& Requests}}. In \bibinfo{booktitle}{\emph{{Proceedings of the 31. ACM International Conference on Multimedia, MM 2023}}}. ACM, \bibinfo{publisher}{ACM}, \bibinfo{address}{Ottawa, Canada}.
\newblock
\newblock
\shownote{5 pages}.


\bibitem[Serdyuk et~al\mbox{.}(2018)]%
        {serdyuk2018towards}
\bibfield{author}{\bibinfo{person}{Dmitriy Serdyuk}, \bibinfo{person}{Yongqiang Wang}, \bibinfo{person}{Christian Fuegen}, \bibinfo{person}{Anuj Kumar}, \bibinfo{person}{Baiyang Liu}, {and} \bibinfo{person}{Yoshua Bengio}.} \bibinfo{year}{2018}\natexlab{}.
\newblock \showarticletitle{Towards end-to-end spoken language understanding}. In \bibinfo{booktitle}{\emph{2018 IEEE International Conference on Acoustics, Speech and Signal Processing (ICASSP)}}. IEEE, \bibinfo{pages}{5754--5758}.
\newblock


\bibitem[Sheikh et~al\mbox{.}(2022)]%
        {sheikh2022introducing}
\bibfield{author}{\bibinfo{person}{Shakeel~Ahmad Sheikh}, \bibinfo{person}{Md Sahidullah}, \bibinfo{person}{Fabrice Hirsch}, {and} \bibinfo{person}{Slim Ouni}.} \bibinfo{year}{2022}\natexlab{}.
\newblock \showarticletitle{Introducing ECAPA-TDNN and Wav2Vec2. 0 embeddings to stuttering detection}.
\newblock \bibinfo{journal}{\emph{arXiv preprint arXiv:2204.01564}} (\bibinfo{year}{2022}).
\newblock


\bibitem[Shor and Venugopalan(2022)]%
        {shor22_interspeech}
\bibfield{author}{\bibinfo{person}{Joel Shor} {and} \bibinfo{person}{Subhashini Venugopalan}.} \bibinfo{year}{2022}\natexlab{}.
\newblock \showarticletitle{{TRILLsson: Distilled Universal Paralinguistic Speech Representations}}. In \bibinfo{booktitle}{\emph{Interspeech 2022}}. \bibinfo{pages}{356--360}.
\newblock
\urldef\tempurl%
\url{https://doi.org/10.21437/Interspeech.2022-118}
\showDOI{\tempurl}


\bibitem[Sun et~al\mbox{.}(2019)]%
        {sun2018functional}
\bibfield{author}{\bibinfo{person}{Shengyang Sun}, \bibinfo{person}{Guodong Zhang}, \bibinfo{person}{Jiaxin Shi}, {and} \bibinfo{person}{Roger Grosse}.} \bibinfo{year}{2019}\natexlab{}.
\newblock \showarticletitle{{FUNCTIONAL} {VARIATIONAL} {BAYESIAN} {NEURAL} {NETWORKS}}. In \bibinfo{booktitle}{\emph{International Conference on Learning Representations}}.
\newblock
\urldef\tempurl%
\url{https://openreview.net/forum?id=rkxacs0qY7}
\showURL{%
\tempurl}


\bibitem[Vaessen and Van~Leeuwen(2022)]%
        {vaessen2022fine}
\bibfield{author}{\bibinfo{person}{Nik Vaessen} {and} \bibinfo{person}{David~A Van~Leeuwen}.} \bibinfo{year}{2022}\natexlab{}.
\newblock \showarticletitle{Fine-tuning wav2vec2 for speaker recognition}. In \bibinfo{booktitle}{\emph{ICASSP 2022-2022 IEEE International Conference on Acoustics, Speech and Signal Processing (ICASSP)}}. IEEE, \bibinfo{pages}{7967--7971}.
\newblock


\end{thebibliography}


\end{document}